\documentclass[twocolumn,showpacs,preprintnumbers,amsmath,amssymb]{revtex4}
\usepackage{graphicx}
\usepackage{dcolumn}
\usepackage{bm}
\usepackage{bbold}
\usepackage{booktabs}

\begin{document}

\def\prb{Phys. Rev. B}
\def\prl{Phys. Rev. Lett.}
\def\pla{Phys. Lett. A}
\def\pr{Phys. Rev.}
\def\be{\begin{equation}}
\def\ee{\end{equation}}
\def\ba{\begin{eqnarray}}
\def\ea{\end{eqnarray}}

\title{Universal Scaling of the N\'eel Temperature of Near-Quantum-Critical\\
Quasi-Two-Dimensional Heisenberg Antiferromagnets}

\author{D.~X.~Yao and A.~W.~Sandvik}

\affiliation{Department of Physics, Boston University, 590 Commonwealth Avenue, 
Boston, Massachusetts 02215, USA}

\date{June 13, 2006}

\begin{abstract}
We use a quantum Monte Carlo method to calculate the N\'eel temperature $T_{\rm N}$ of weakly 
coupled $S=1/2$ Heisenberg antiferromagnetic layers consisting of coupled ladders. This system can be 
tuned to different two-dimensional scaling regimes for $T > T_{\rm N}$. In a single-layer mean-field 
theory, $\chi_s^{2D}(T_{\rm N})=(z_2J')^{-1}$, where $\chi_s^{2D}$ is the exact staggered susceptibility 
of an isolated layer, $J'$ the inter-layer coupling, and $z_2=2$ the layer coordination number. 
With a renormalized $z_2$, $z_2 \to k_2z$, we find that this relationship applies not only in the 
renormalized-classical regime, as shown previously, but also in the quantum-critical regime and part 
of the quantum-disordered regime. The renormalization is nearly constant; $k_2 \approx 0.65 - 0.70$.
We also study other universal scaling functions.
\end{abstract}


\maketitle

Antiferromagnets with effectively low dimensionality, consisting of weakly coupled chains (quasi-1D) 
or layers (quasi-2D), offer unique opportunities to study quantum mechanical collective behavior. 
A multitude of quasi-1D and quasi-2D antiferromagnetic compounds have been discovered, or deliberately 
designed, and they exhibit a wide range of ordered and disordred phases. At the same time, new and 
improved experimental techniques enable increasingly sophisticated studies of their properties. 
It is thus possible to test in detail microscopic quantum spin models and theoretical quantum many-body 
concepts, such as quantum-critical scaling \cite{sachdevbook}. Motivation for studying these systems 
often come from phenomena directly associated with low dimensionality. Real materials, however, almost 
always have some three-dimensional (3D) couplings that cannot be completely ignored at low temperatures. 
The ways in which these couplings change the physics, e.g., leading to phase transitions or dimensional 
cross-overs \cite{liu}, are also governed by the physics of the 1D or 2D units. Studies of 3D effects 
can therefore also provide important insights. 

In this Letter we investigate the finite-temperature N\'eel transition temperature $T_{\rm N}$ of a 
quasi-2D $S=1/2$ Heisenberg model consisting of layers of coupled ladders. In the absence of inter-layer 
couplings, the Mermin-Wagner theorem dictates that the system can have long-range order only at $T=0$. 
The 2D Heisenberg model with spatially isotropic nearest-neighbor couplings $J$ has an ordered ground 
state \cite{reger,chn}. At low temperatures, in the renormalized classical (RC) regime, its spin 
correlation length is exponentially divergent \cite{chn}. Systems with a coupling pattern favoring 
formation of nearest-neighbor singlets \cite{singh}, e.g., coupled two-leg ladders \cite{takayama01}, 
can be tuned through a quantum phase transition into a quantum disordered (QD) state. This $T=0$ 
transition and its associated $T>0$ quantum-critical (QC) scaling regime have been studied in detail, 
using field-theoretical approaches \cite{chubukov} and quantum Monte Carlo (QMC) simulations 
\cite{bilayqmc,troyer,takayama01}. Some QMC studies of the N\'eel transition in systems of weakly 
coupled spatially isotropic layers have also been reported. Sengupta et al.~studied $T_{\rm N}$ 
and the dimensional cross-over in the specific heat \cite{sandvik03a}. Yasuda et al. studied  $T_{\rm N}$ 
more systematically at very small ratios $\alpha=J'/J$ of the inter- and intra-layer couplings, 
for $S=1/2$ as well as higher spins \cite{yasuda05a}. 

We here carry out QMC calculations of the coupled-ladder system to test three recently proposed scaling 
functions relating $T_{\rm N}$ and various 2D and 3D staggered susceptibilities \cite{mudry06a,praz06a}. 
We focus in particular on inter-ladder couplings for which the system is near-quantum-critical in the 
absence of inter-layer couplings. Our main finding is that two of the scaling functions are almost 
constant and do not change appreciably between the RC and QC regimes. In particular, the coordination 
number renormalization introduced by Yasuda et al.~changes from $k_2 \approx 0.65$ in the RC regime 
\cite{yasuda05a} to $k_2\approx 0.68$ in the QC regime. This minute change implies that the N\'eel
ordering takes place almost exactly at the same temperature at which the planes start to correlate
appreciably, and that these correlations are almost completely governed by the magnitude of the
static staggered susceptibility of the planes. The nature of the fluctuations, RC or QC (for which 
the dynamic susceptibilities are completely different \cite{chubukov}), does not play a major role.

In a single-layer mean-field theory (also referred to as RPA) \cite{scalapino}, the 3D couplings of
a layer $l=0$ are taken into account by a static staggered magnetic field arising from the ordered 
moments of the two adjacent planes $l = \pm 1$. The self-consistent N\'eel temperature is then 
obtained by solving the equation
\begin{equation}
\chi_s^{2D}(T_N)=(z_2 \alpha)^{-1},
\label{tnmeanfield}
\end{equation}
where $z=2$ is the layer coordination number and $\chi_s^{2D}$ is the staggered 
susceptibility of a single layer, the $T$ dependence of which is known from studies of the
quantum nonlinear $\sigma$ model. In the RC regime, $T < 4\pi \rho_s$ \cite{chn}, 
\begin{equation}
\chi_s^{2D}(T)\propto T{\rm e}^{4\pi \rho_s/T},
\label{chisrc}
\end{equation}
where $\rho_s$ is the spin stiffness. Using numerically exact QMC results for $\chi_s^{2D}$ and $T_{\rm N}$, 
Yasuda et al.~\cite{yasuda05a} found that the mean-field expression (\ref{tnmeanfield}) accurately  
captures the $\alpha \ll 1$ dependence of $T_{\rm N}$, if $z$ is replaced by a renormalized 
coordination number $z_2k_2$;
\begin{equation}
\chi_s^{2D}(T_{\rm N})=(k_2z_2\alpha)^{-1}.
\end{equation}
Moreover, the renormalization, $k_2 \approx 0.65$, was found to be independent on the spin $S$. This 
intriguing result prompted Hastings and Mudry to carry out a detailed renormalization group (RG) 
study of the anisotropic $O(N)$ non-linear sigma model \cite{mudry06a}. Instead of a constant 
coordination number renormalization, they argued that the quantity 
\begin{equation}
F_1=  (k_2z_2)^{-1}= \alpha\chi_s^{2D}
\label{f1}
\end{equation}
is a universal function of $x=c[T_{\rm N}\xi^{\rm 2D}(T_{\rm N})]^{-1}$ when $\alpha \ll 1$. Here  
$\xi^{2D}$ is the correlation length of a single isolated layer and $c$ the spinwave velocity. They 
concluded that the reason for the near constant $k_2$ is that the single layer is in the RC regime 
for all $S$ at low $T$, whence  $x$ is exponentially small and $F_1( x\to 0)$ is constant. In the QC and QD 
regimes $F_1$ should approach other constant values. A quantity involving the susceptibility 
$\chi({\bf Q})$ of the full 3D system at wave-vector ${\bf Q}=(\pi,\pi,0)$
was also introduced
\begin{equation}
F_2=\alpha\chi(\pi,\pi,0).
\label{f2}
\end{equation}
To leading order in an $1/N$ approximation, Praz et al.~\cite{praz06a} 
found that $F_2 =1/4$ in all regimes when $\alpha \to 0$. This prediction should be easier to test 
experimentally because it involves only properties of the actual quasi-2D system. They also proposed 
a third universal
quantity
\begin{equation}
F_3=\alpha S(\pi,\pi,0)T_N^{-1},
\label{f3}
\end{equation}
where $S({\bf Q})$ is the static spin structure factor. This function was shown to distinguish between
the RC, QC, and QD regimes already at the $N=\infty$ level.

Since the $\alpha \to 0$ values of $F_1,F_2$, and $F_3$ were evaluated at the $N=\infty$ level 
or including only order-$1/N$ corrections, significant higher-order corrections to these results 
were expected \cite{praz06a}. Unbiased numerical results would therefore be useful. The previous QMC 
results by Yasuda et al.~for $k_2=[2F(x\to 0)]^{-1}$ in the RC regime, $k_2 \approx 0.65$ \cite{yasuda05a}, 
falls between the $N=\infty$ and $1/N$ values; $k_2=1/2$ and $1.01$ \cite{praz06a}. $F_2$ and $F_3$  
have not yet been calculated in the RC regime, and none of the predictions have been tested 
against numerical results in the QC and QD regimes. 

The universal constants could be very useful for
extracting inter-layer couplings experimentally. This motivates us to carry out large-scale QMC simulations 
of the coupled-ladder system, where the individual layers can be tuned through a quantum-critical 
point. We can then obtain numerical results for $F_1$-$F_3$ in all three 2D temperature regimes. 

\begin{figure}
\includegraphics[width=4.5cm]{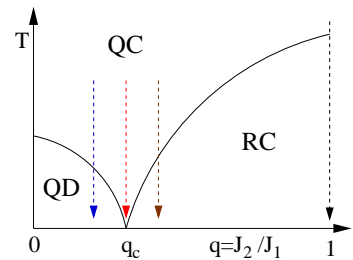}
\caption{Finite-temperature regimes for a 2D layer of coupled ladders with inter-ladder 
coupling $q$. The curves indicate cross-over temperatures. We study $T>0$ properties of  
a quasi-2D system (coupled layers) with  $q$ at and close to the 2D critical point $q_c$
and at the isotropic point $q=1$.}
\label{phase}
\end{figure}

The hamiltonian for coupled Heisenberg layers consisting of two-leg ladders is
\begin{equation}
H= J_1\sum_{\langle i,j\rangle_1}\mathbf{S}_i \cdot \mathbf{S}_j +J_2 \sum_{\langle i,j\rangle_2}
\mathbf{S}_i \cdot \mathbf{S}_j+J_3 \sum_{\langle i,j\rangle_3} \mathbf{S}_i \cdot
\mathbf{S}_j,      
\label{hamilton}
\end{equation}
where $\langle i,j\rangle_1$ denotes a pair of nearest-neighbor spins in the same ladder,
$\langle i,j\rangle_2$ in different ladders of the same layer, and $\langle i,j\rangle_3$
in adjacent layers. We define the coupling ratios $q=J_2/J_1$ and $\alpha = J_3/J_1$. 

The quantum phase transition of the single layer ($\alpha=0$) has been studied by Matsumoto 
et al.~\cite{takayama01}. The critical coupling $q_c=0.31407(5)$. The $T>0$  cross-overs 
for an isolated layer are shown schematically in Fig.~\ref{phase}. In the limit 
$\alpha \to 0$, the quantum-critical coupling of the quasi-2D system approaches $q_c$ 
of the single layer. In our study we focus on values of $q$ close to the 2D quantum-critical 
point,  choosing $q=0.25, 0.30, 0.31407=q_c$, and $0.33$. We also consider the previously 
studied case $q=1$ \cite{yasuda05a}, to calculate also $F_2$ and $F_3$ deep inside the RC regime. 
We have obtained results for $\alpha$ in the range $10^{-3}$ to $1$.

We use the stochastic series expansion (SSE) QMC method \cite{sandvik99} to study periodic lattices with 
$L_xL_yL_z$ spins, with $L_x=L_y=L$ up to $128$. To take into account, at least partially, the fact 
that $\xi_{x,y} \gg \xi_{z}$ when $\alpha \ll 1$, we use aspect ratios $L/L_z$ up to $16$. To determine 
the N\'eel temperature, we use the finite-size scaling of the spin stiffness constants 
$\rho_s^\mu$ in the three different directions, $\mu=x,y,z$, of the spatially anisotropic lattice. 
This approach was previously taken in Ref.~\cite{sandvik03a}. For fixed aspect ratio, the stiffness 
at $T_N$ should scale as $L^{-1}$. We thus locate the point at which $L\rho_\mu(T)$ becomes asymptotically 
size-independent (extrapolating crossing points for different size $L$ to $L \to \infty$). 
For $q=1$, we use $T_N$ from Ref.~\cite{yasuda05a}. For the calculations of the 2D staggered 
susceptibility $\chi_s^{2D}(T)$ we have used $L$ up to $800$.

\begin{figure}
\includegraphics[width=7cm, clip]{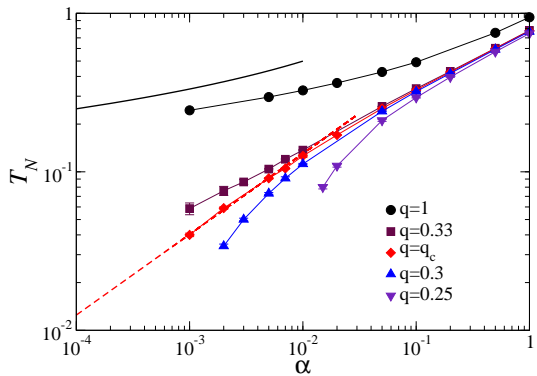}
\caption{N\'eel temperature as a function of inter-layer coupling at different inter-ladder 
couplings $q$. The $q=1$ results are from Ref.~\cite{yasuda05a}. The solid (black) and dashed 
(red) lines shows the expected  $\alpha\to 0$ forms in the RC and QC regimes.} 
\label{tn}
\end{figure}

The $\alpha$ dependence of the N\'eel temperature for the different $q$ values is 
shown on a log-log scale in Fig.~\ref{tn}. When $q < q_c$, there is a minimum value $\alpha_c$
of the inter-layer coupling below which the system cannot order---$\alpha_c(q<q_c)$ is the line
of 3D quantum-critical points and $\alpha_c(q_c)=q_c$. In our results for $q>q_c$ we see a down-turn 
of $T_{\rm N}$ as $\alpha$ decreases, reflecting the 3D quantum-critical point. From our limited 
low-$T_{\rm N}$ data we can only roughly extract two points on the critical line; $\alpha_c (q=0.30) 
\approx 0.001$ and $\alpha_c (q=0.25) \approx 0.006$. 

For $q > q_c$ we can solve the mean-field equation (\ref{tnmeanfield}) with the RC form (\ref{chisrc}) 
of the correlation length, giving, to leading order in $\alpha$,
\begin{equation}
T_N(\alpha) \propto -[\ln(\alpha)]^{-1}.
\end{equation}
As shown in Fig.~\ref{tn}, for $q=1$ this form does not yet apply at $\alpha=10^{-3}$, but it
should be the correct form for $\alpha \to 0$. Yasuda et al.~presented an empirical formula 
that works well also at higher $\alpha$ \cite{yasuda05a}. For $q=0.33$ we should also approach 
the RC form when $\alpha \to 0$, but here we instead observe an almost perfect power law in the 
whole range of $\alpha \ge 10^{-3}$. However, the exponent is not the one expected in the 
QC scaling regime (discussed further below), and we expect the behavior to eventually cross over to 
the log form.

In the QC regime, the 2D staggered susceptibility takes an asymptotic $T \to 0$ power-law form, 
\begin{equation}
\chi_s^{2D}(T)\propto T^{-2+\eta},
\label{xs_qc}
\end{equation}
where $\eta \approx 0.038$ \cite{compostrini} is the correlation function exponent of the
3D $O(3)$ Universality class. Using this form in the mean-field equation (\ref{k2}) we get a
corresponding power-law behavior of $T_N$;
\begin{equation}
T_N(\alpha) \propto \alpha^{1/(2-\eta)}.
\label{qctc}
\end{equation}
In Fig.~\ref{tn} we can see that this quasi-2D quantum-critical form accurately describes the 
results for $q=q_c$ below $\alpha \approx 10^{-2}$. For larger $\alpha$, $T_{\rm N}$ is still
in the high-temperature regime where the behavior is influenced by non-universal lattice effects
\cite{chn,chubukov}. The two $q$ values close to $q_c$, for which the reduced coupling 
$|g_-g_c|/g_c \approx 0.05$, are already too far from the critical point to observe any distinct 
(asymptotic-form) QC behavior before the cross-overs occur. 

\begin{figure}
\includegraphics[width=7cm, clip]{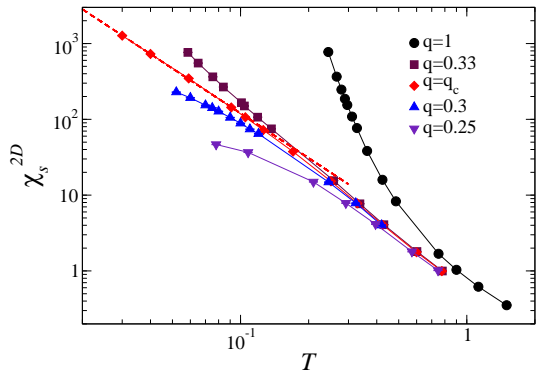}
\caption{Staggered susceptibility $\chi_s^{2D}$ vs. temperature for 2D
systems composed of coupled two-leg ladders with different inter-ladder
coupling ratios $q$. The dashed line shows the asymptotic QC power-law behavior,
Eq.~(\ref{qctc}).}
\label{xs}   
\end{figure}

\begin{figure}
\includegraphics[width=5.75cm, clip]{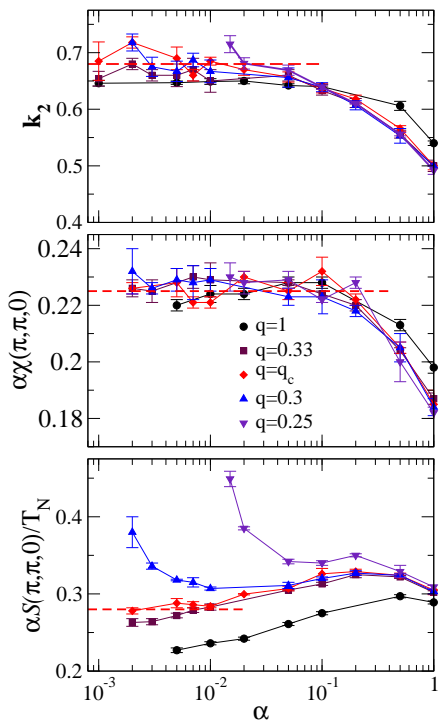}
\caption{Inter-layer coupling dependences of the quantities (\ref{f1}), (\ref{f2}), and (\ref{f3}) 
for different inter-ladder couplings $q$. The dashed lines show our extracted constant values
for the QC regime.}
\label{scaling}
\end{figure}

Following Ref.~\cite{yasuda05a}, we study the coordination number renormalization
\begin{equation}
k_2(\alpha)=[2\alpha \chi_s^{2D}(T_N)]^{-1} = (2F_1)^{-1}.
\label{k2}
\end{equation}
In Fig.~\ref{xs} we show our QMC results for the staggered susceptibility of the isolated
2D layers. Using these results and the $T_{\rm N}$ data shown in Fig.~\ref{tn}, we obtain the 
results for $k_2$ shown in the upper panel of Fig.~\ref{scaling}. For $q=1$, Yasuda et al.~found
$k_2 \approx 0.65$ for $\alpha < 0.1$ \cite{yasuda05a}. We here show $q=1$ results obtained
with their listed $T_{\rm N}$ values and our own results for $\chi_s^{\rm 2D}$. The resulting
$k_2$ agree with the previous results. Surprisingly, we hardly see any change 
in $k_2$ when going to the near-critical systems, except some small differences when $\alpha > 0.1$. 
At lower $\alpha$, $k_2$ is only a few percent larger for $q\approx q_c$ than at 
$q=1$; $k_2(q_c) \approx 0.68$. Even for our $q < q_c$ points, $k_2$ remains close to this value, 
even though $T_{\rm N}$ is seen crossing over into QD behavior in Fig.~\ref{tn}. For the lowest 
$\alpha$ considered for $q=0.25$ and $0.30$  we see a slight increase in $k_2$, but the effect 
is barely statistically significant. Note that for $q<q_c$, $k_2$ is not defined for 
$\alpha < \alpha_c(q)$. 

In the RG study by Praz et al.~\cite{praz06a}, different expressions for $F_1$ ($k_2$) were 
obtained in saddle-point approximations for the RC, QC, and QD regimes. No numerical values 
were given, however, except for $k_2=1/2$ in the RC regime. Corrections to the constant behavior was 
expected (and calculated to order $1/N$ in the case of the RC regime, then giving $k_2=1.01$). 
Furthermore, significantly different constants were expected for the three regimes. The near 
constant $k_2 \approx 0.65 - 0.70$ we find here for such a wide range of $q$ values, spanning
all three temperature regimes, is thus quite remarkable.

We now turn to the second scaling function, Eq.~(\ref{f2}). As seen in the middle panel of 
Fig.~\ref{scaling}, we obtain an almost constant $F_2=\alpha\chi(\pi,\pi,0)\approx 0.22-0.23$ for all 
$q$ and for a wide range of $\alpha$. In the $N=\infty$ approximation, $F_2=1/4$ in all temperature 
regimes \cite{praz06a}, remarkably close to what we find here. However, also in this case the actual 
values in the RC, QC, and QD regimes were expected to differ markedly once $1/N$ and higher corrections 
are included.

The third scaling function, Eq.~(\ref{f3}), distinguishes between the RC, QC, and QD regimes already
at the $N=\infty$ level \cite{praz06a}. In the RC regime $S({\bf Q})=T\chi({\bf Q})$ to leading 
order \cite{chn} and thus $F_3 = F_2 =1/4$. Our results for the RC regimes ($q=1$), shown in 
the bottom panel of Fig.~\ref{scaling}, are slightly lower than the predicted value for 
$\alpha < 0.1$. There is also still a decreasing trend as $\alpha$ decreases and we cannot
reliably extract the asymptotic constant RC value (for $\alpha=10^{-3}$ our calculations
for $q=1$ are not completely size converged and we therefore do not show them here). For $q=0.33$, 
which also should give RC behavior for $\alpha \to 0$, 
the results are still quite far from the $q=1$ curve for all $\alpha$, but the 
decreasing trend is consistent with the same asymptotic value. For $F_3$ we also see clear 
differences in behavior in the three regimes. There are distinct cross-overs from QC to RC 
or QD behavior. The results for $q=q_c,0.30$, and $0.33$ all fall on the same universal QC curve 
for $\alpha$ down to $\approx 0.05$, below which the $q=0.25$ curve splits off. The $q=q_c$ and $0.3$ 
curves coincide to even lower $\alpha$. The asymptotic value at $q_c$ is $\approx 0.27$.
For $q<q_c$ we expect a divergence at $\alpha_c(q)$, as $S(\pi,\pi,0)$ must converge to a constant
when $q<q_c$ and $T_{\rm N} \to 0$. We see clear signs of this divergence. 

In conclusion, we have presented results for the quantities $\alpha \chi_s^{2D}\equiv 1/2k_2$,
$\alpha \chi(\pi,\pi,0)$, and $\alpha S(\pi,\pi,0)/T$, at $T=T_{\rm N}$ for a quasi-2D system
of coupled ladder. For weak inter-layer
coupling, $\alpha \to 0$, these quantities have been predicted to take different universal 
constant values in the RC, QC, and QD regimes \cite{mudry06a,praz06a}. We have investigated the
dependence on $\alpha$ for $10^{-3} \le \alpha \le 1$, with the goal of extracting the constants
and investigate the $\alpha > 0$ corrections. We find a remarkably stable value of the coordination 
number renormalization $k_2$ and $\alpha \chi(\pi,\pi,0)$: For $\alpha < 0.1$ they are almost 
independent on $\alpha$ and do not change appreciably between the RC, QC, and QD regimes. 
Significant differences in the three regimes were anticipated based on the previous RG 
study of the nonlinear $\sigma$ model by Praz et al.~\cite{praz06a}. Only in $\alpha S(\pi,\pi,0)$ 
do we see distinct differences. It would be useful to extend the calculations to still lower 
inter-layer couplings, but reaching significantly below $\alpha=10^{-3}$ with QMC requires 
prohibitively large lattices. 

The almost constant $F_1$ and $F_2$ imply that the correlations between layers is predominantly 
governed by the magnitude of the static staggered susceptibility of the layers. The range of 
temperatures for which the system is 3D critical is almost negligible, regardless of the nature 
of the 2D fluctuations---RC or QC---that initially lead to correlations between the layers.

We would like to thank D.~K.~Campbell, M.~B.~Hastings, C.~Mudry,  and A.~Praz for useful 
discussion. This work was supported by the NSF under grant No.~DMR-0513930 and by Boston University.

\null

\end{document}